\documentclass[aps,preprint,preprintnumbers,amsmath,amssymb,amsfonts,groupedaddress,superscriptaddress,showpacs,showkeys,floatfix]{revtex4-2}

\usepackage{graphicx}
\usepackage{subfigure}
\usepackage{bm}% bold math
\usepackage{amsmath}
\usepackage{color}
\usepackage{array}
\usepackage{CJK}

\newcommand{\bmath}{\begin{mathletters}}
\newcommand{\emath}{\end{mathletters}}
\newcommand{\be}{\begin{eqnarray}}
\newcommand{\ee}{\end{eqnarray}}
\newcommand{\ba}{\begin{array}}
\newcommand{\ea}{\end{array}}

\newcommand{\ra}{\rangle}

\newcommand{\no}{\nonumber}

\newcommand{\pr}{\prime}

\newcommand{\Tr} {\mathrm{Tr}}

\begin{document}
\title{Optimization of Controlled-$Z$ Gate with Data-Driven Gradient Ascent Pulse Engineering in a Superconducting Qubit System}

\author{Zhiwen Zong}
   \thanks{These authors have contributed equally to this work.}
 \affiliation{Zhejiang Province Key Laboratory of Quantum Technology and Device, Department of Physics, Zhejiang University, Hangzhou, 310027, China}
 \author{Zhenhai Sun}
    \thanks{These authors have contributed equally to this work.}
 \affiliation{Zhejiang Province Key Laboratory of Quantum Technology and Device, Department of Physics, Zhejiang University, Hangzhou, 310027, China}
   \author{Zhangjingzi Dong}
 \affiliation{Zhejiang Province Key Laboratory of Quantum Technology and Device, Department of Physics, Zhejiang University, Hangzhou, 310027, China}
   \author{Chongxin Run}
 \affiliation{Zhejiang Province Key Laboratory of Quantum Technology and Device, Department of Physics, Zhejiang University, Hangzhou, 310027, China}
\author{Liang Xiang}
 \affiliation{Zhejiang Province Key Laboratory of Quantum Technology and Device, Department of Physics, Zhejiang University, Hangzhou, 310027, China}
  \author{Ze Zhan}
 \affiliation{Zhejiang Province Key Laboratory of Quantum Technology and Device, Department of Physics, Zhejiang University, Hangzhou, 310027, China}
  \author{Qianlong Wang}
 \affiliation{Zhejiang Province Key Laboratory of Quantum Technology and Device, Department of Physics, Zhejiang University, Hangzhou, 310027, China}
   \author{Ying Fei}
 \affiliation{Zhejiang Province Key Laboratory of Quantum Technology and Device, Department of Physics, Zhejiang University, Hangzhou, 310027, China}
   \author{Yaozu Wu}
 \affiliation{Zhejiang Province Key Laboratory of Quantum Technology and Device, Department of Physics, Zhejiang University, Hangzhou, 310027, China}
   \author{Wenyan Jin}
 \affiliation{Zhejiang Province Key Laboratory of Quantum Technology and Device, Department of Physics, Zhejiang University, Hangzhou, 310027, China}
    \author{Cong Xiao}
 \affiliation{Zhejiang Province Key Laboratory of Quantum Technology and Device, Department of Physics, Zhejiang University, Hangzhou, 310027, China}
 \author{Zhilong Jia}
 \affiliation{Key Laboratory of Quantum Information, University of Science and Technology of China, Hefei, 230026, China}
 \author{Peng Duan}
 \affiliation{Key Laboratory of Quantum Information, University of Science and Technology of China, Hefei, 230026, China}
 \author{Jianlan Wu }
 \email{jianlanwu@zju.edu.cn}
 \affiliation{Zhejiang Province Key Laboratory of Quantum Technology and Device, Department of Physics, Zhejiang University, Hangzhou, 310027, China}
 \author{Yi Yin}
 \email{yiyin@zju.edu.cn}
 \affiliation{Zhejiang Province Key Laboratory of Quantum Technology and Device, Department of Physics, Zhejiang University, Hangzhou, 310027, China}
 \author{Guoping Guo}
 \email{gpguo@ustc.edu.cn}
 \affiliation{Key Laboratory of Quantum Information, University of Science and Technology of China, Hefei, 230026, China}
 \affiliation{Origin Quantum Computing, Hefei, 230026, China}

\begin{abstract}
The experimental optimization of a two-qubit controlled-$Z$ (CZ) gate is realized following two different data-driven gradient ascent pulse engineering (GRAPE) protocols
in the aim of optimizing the gate operator and the output quantum state, respectively. For both GRAPE protocols, the key computation of gradients utilizes mixed information
of the input $Z$-control pulse and the experimental measurement. With an imperfect initial pulse in a flattop waveform, our experimental implementation shows that the
CZ gate is quickly improved and the gate fidelities subject to the two optimized pulses are around 99\%.
Our experimental study confirms the applicability of the data-driven GRAPE protocols in the problem of the gate optimization.

\end{abstract}

\maketitle

\section{Introduction}

The realization of high-fidelity quantum gates is essential in quantum computation and quantum simulation~\cite{ChuangBook}.
As an important one in the group of fundamental quantum gates,
the two-qubit controlled-NOT (CNOT) gate can be experimentally created by the combination of a two-qubit controlled-$Z$ (CZ) gate and
two single-qubit gates~\cite{BarendsNat14,BarendsPRL19,ChowPRL11,YanFeiPRApp18,CaldwellPRApp18}.
The recent advancements in technology have allowed precise control and measurement of quantum devices.
The superconducting qubit system has reached $<\!1\%$ errors below the fault-tolerant threshold of surface code quantum
computing~\cite{FowlerPRA12,BarendsNat14,KellyPRL14}. In our previous study of the CZ gate,
the gate fidelity is $\sim94\%$ for a shortcut-to-adiabaticity (STA) pulse~\cite{wthPRAPP19}.
Although such an external pulse with an analytic form is experimentally available~\cite{DiCarloNat09,MartinisPRA14},
the state-of-the-art high-fidelity gate still needs optimization algorithms due to unavoidable control
distortion. In a previous work by Martinis and his coworkers, the fidelity of the CZ gate reaches $>99\%$ under an optimal fast adiabatic pulse~\cite{KellyPRL14}.
This optimization is realized by a randomized benchmarking (RB) based Nelder-Mead learning algorithm~\cite{KellyPRL14,NelderCJ65}.
In a RB experiment, the statistical average of the ground-state population over sequences of random Clifford gates
is utilized to identify the fidelity of a specific quantum gate~\cite{ChowPRL09,MagesanPRL11,IBMPRL12}.
Following a test-and-trial strategy, the Nelder-Mead algorithm searches the parameter space
for an optimization point. Despite its simple implementation,
this algorithm is fundamentally slow
since the gate fidelity is statistically determined and cannot be described as a simple functional of the external control pulse.

Instead, we can apply a gradient-based optimization since the gate operator is
equivalent to a time evolution operator fully dependent on the control pulse.
Through the time discretization, the control pulse is changed to a sequence of pulse amplitudes
at various time points and the derivative of the gate operator over each pulse amplitude can be numerically
calculated, which leads to a gradient ascent pulse engineering (GRAPE) algorithm~\cite{KhanejaJMR94}.
In comparison with the Nelder-Mead algorithm, the GRAPE algorithm yields a much faster search 
due to the guidance of gradient vectors and a great flexibility is allowed
in the dimensionality of the parameter space.

In the original design of the GRAPE algorithm~\cite{KhanejaJMR94,MotzoiPRA11}, the numerical
calculation of the gradient vector needs an accurate theoretical description of
quantum dynamics, which is not always available in real experiments due to systematic errors.
A hybrid approach with information of the experimental measurement
can partially circumvent this difficulty~\cite{JudsonPRL92,BrifNJP10,EggerPRL14}. Following the feedback-control technique,
various data-driven GRAPE protocols have been proposed and implemented in
the state preparation and the gate optimization~\cite{ChangpuSunPRL17,DaweiLuNPJ17,LaflammePRA18,RebingWuPRA18}.
For the CZ gate, the gate operator can be fitted by
the Powell method over the experimental measurement of the quantum process tomography (QPT)~\cite{MohseniPRA08,YamamotoPRB10,ChowPRL12},
which collects the data of the quantum state tomography (QST) generated from 36 initial states.
Despite its intrinsic advantages, the data-driven GRAPE protocol by optimizing the gate operator still
carries a heavy experimental burden. On the other hand, not all the initial states in the QPT measurement
are equally important in the evaluation of the CZ gate. We can select one or few relevant
initial states and optimize the control pulse for the best output density matrices.
This state optimization provides an alternative approach of the gate optimization.

The rest of this paper is organized as follows. In Secs.~\ref{sec2} and \ref{sec3}, we
provides the data-driven GRAPE protocol based on the optimization of the CZ gate operator
and presents the results of experimental implementation in the system of
two superconducting X-shaped transmon qubits. In Secs.~\ref{sec4} and \ref{sec5}, we
provides the GRAPE protocol based on the optimization of the density matrix
and presents the experimental results. In Sec.~\ref{sec6}, we summary our experimental study.

\begin{figure}[tp]
\centering
 \includegraphics[width=0.65\columnwidth]{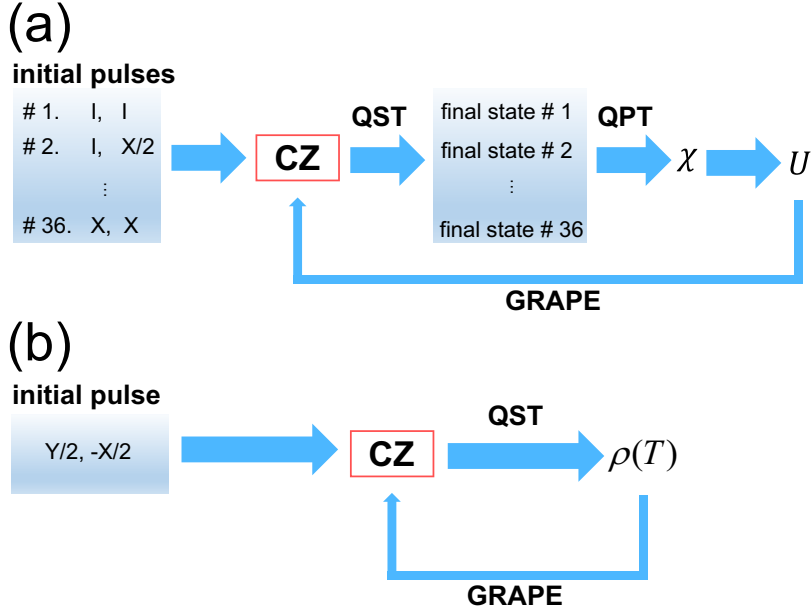}
\caption{Schematic diagrams of two data-driven GRAPE protocols implemented in our experiment.
These two protocols are designed under the optimization of (a) the fitted gate operator $U$
and (b) an output density matrix $\rho(T)$, respectively.}
\label{fig_n01}
\end{figure}

\section{Data-Drive GRAPE Protocol I}
\label{sec2}

In this section, we provide the theoretical description of our first
data-driven GRAPE protocol for the realization of the CZ gate, similar
to the design in Ref.~\cite{RebingWuPRA18}. A schematic diagram of
this protocol is shown in Fig.~\ref{fig_n01}(a).
The Hamiltonian of a two-qubit ($A$ and $B$) system is written as
\be
H_{0} = H_A  + H_B + H_\mathrm{int},
\label{eq_001}
\ee
where $H_{A}$ and $H_B$ are two single-qubit Hamiltonians,
and $H_\mathrm{int}$ is the interaction between two qubits. Since our CZ gate is assisted by the second excited
state of one qubit~\cite{wthPRAPP19}, a three-level model is
considered in the single-qubit Hamiltonian,
\be
 H_{i=A,B} = \omega_i|1_i\rangle\langle1_i| + \left(2\omega_i+\Delta_i)\right|2_i\rangle\langle2_i|.
\label{eq_002}
\ee
For each qubit ($i=A, B$), $\omega_i$ and $\Delta_i$ are its resonant
frequency and anharmonicity parameter, respectively. The reduced Planck
constant $\hbar$ is set to be unity throughout this paper. In our
experiment, the frequency shift is $\delta\omega/2\pi=(\omega_A-\omega_B)/2\pi= 539.0$~MHz
and the two anharmonicity parameters are $\Delta_A/2\pi=-242.1$ MHz
and $\Delta_B/2\pi=-258.8$ MHz. The interaction term is written as
\be
 H_\mathrm{int} = g (a_A^\dagger  a_B + a_A a_B^\dagger),
\label{eq_003}
\ee
where $a_{i=A,B}=\sum_{j=1}^{2}\sqrt{j}|(j-1)_i\rangle\langle j_i|$
and $a_{i=A,B}^\dagger=\sum_{j=0}^{1}\sqrt{j+1}|(j+1)_i\rangle\langle j_i|$
are the lowering and raising operators, respectively.
In our experiment, the coupling strength is $g/2\pi=9.1$ MHz.

Due to the condition of a weak interaction
($g\ll \delta\omega, |\Delta_A|, |\Delta_B|$), the population exchange between
two qubits is usually negligible, but a $Z$-control pulse can
tune the energy levels and create an inter-qubit resonance. In our experiment,
the $Z$-pulse $\mu_A(t)$ is applied to qubit $A$, which gives rise to
\be
H_\mathrm{ext}(t) = \mu_A(t) n_{A},
\label{eq_004}
\ee
with the number operator $n_{A}=\sum_{j=0}^{2}j|j_A\rangle\langle j_A|$.
The coupled Hamiltonian, $H_c(t)=H_0(g)+H_\mathrm{ext}(t)$, creates the
resonance between $|2_A0_B\rangle$ and $|1_A1_B\rangle$ under the
pulse amplitude, $\mu_A(t)=\mu_{A; r}=-(\delta\omega+\Delta_A)$.
For conciseness, the notation of an arbitrary state, $|j_Aj^\pr_B\rangle$,
is abbreviated to $|jj^\pr\rangle$ where the first and
second state indices refer to qubits $A$ and $B$, respectively.

In a simplified treatment, the Hilbert space is reduced to
$\{|00\rangle, |10\rangle, |01\rangle, |11\rangle, |20\rangle\}$,
while the coupling only exists between $|11\rangle$ and $|20\rangle$
with the strength $\sqrt{2}g$.
If the energy difference between these two states is precisely tuned to zero
($\mu_A(0<t<T)=\mu_{A; r}$) and the operation time is equal to one period of
the Rabi oscillation ($T=\pi/\sqrt{2}g$), a $\pi$-phase is generated
for $|11\rangle$ and $|20\rangle$. In the five-state
Hilbert space, the time evolution operator is given by
\be
U_c&=&|00\rangle\langle00|+e^{-i\phi_A}|10\rangle\langle10|+e^{-i\phi_B}|01\rangle\langle01|\no \\
&&-e^{-i(\phi_A+\phi_B)}|11\rangle\langle11|-e^{-i(\phi_A+\phi_B)}|20\rangle\langle20|,
\label{eq_005}
\ee
where $\phi_A=\int_0^T [\omega_A+\mu_A(t)]dt$ and $\phi_B=\omega_B T$
are the dynamic phases associated with the first excited states of
qubits $A$ and $B$, respectively. In experiment, these two dynamic
phases can be measured and compensated~\cite{wthPRAPP19}, which is
described by an auxiliary operator,
\be
U^\dagger_d&=&|00\rangle\langle00|+e^{i\phi_A}|10\rangle\langle10|+e^{i\phi_B}|01\rangle\langle01|\no \\
&&+e^{i(\phi_A+\phi_B)}|11\rangle\langle11|+e^{i(\phi_A+\phi_B)}|20\rangle\langle20|.
\label{eq_006}
\ee
This operator can be viewed as a reversed time evolution over a
decoupled Hamiltonian, $H_d(t)=H_0(g=0)+H_\mathrm{ext}(t)$.
The combination of these two operations gives rise to the ideal CZ gate,
\be
U_{\mathrm{CZ}}&=&U^\dagger_dU_c=|00\rangle\langle00|+|10\rangle\langle10|+\no \\
&&|01\rangle\langle01|-|11\rangle\langle11|-|20\rangle\langle20|.
\label{eq_007}
\ee
Note that a standard CZ gate does not involve the evolution
of state $|20\rangle$, which is satisfied in our treatment if
the initial quantum state is inside the subspace of
$\{|00\rangle, |10\rangle, |01\rangle, |11\rangle\}$.

The experimental realization of an ideal square-shaped pulse is difficult due to the bandwidth limitation of the waveform generator.
The residual errors in the control line in general cause that the input pulse experienced by the qubit sample deviates from its theoretical design.
In literature, various approaches have been designed to modify the pulse shape and improve
the gate fidelity~\cite{KellyPRL14,MartinisPRA14,wthPRAPP19}.
In this paper, we apply a data-driven GRAPE method
as follows. The operation time $T$ is discretized into $M$ segments,
each with the same length $\tau=T/M$.
The $Z$-control pulse becomes an amplitude sequence,
i.e., $\mu_A(t)\Rightarrow\{\mu_{A;1}, \mu_{A;2}, \cdots, \mu_{A;M}\}$, 
which leads to $H_{c}(t)\Rightarrow\{H_{c;1}, H_{c; 2}, \cdots, H_{c; M}\}$
and $H_{d}(t)\Rightarrow\{H_{d;1}, H_{d; 2}, \cdots, H_{d; M}\}$.
The two Hamiltonians are given by
$H_{c; m}=H_0(g)+H_\mathrm{ext}(\mu_{A;m})$ and $H_{d; m}=H_0(g=0)+H_\mathrm{ext}(\mu_{A;m})$
at each $m$-th time segment.
The two time evolution operators in Eqs.~(\ref{eq_005}) and (\ref{eq_006})
are expanded into
\be
U_c &=& U_{c; M}U_{c; M-1} \cdots U_{c; 2} U_{c; 1}, \label{eq_008} \\
U^\dagger_d &=& U^\dagger_{d; 1}U^\dagger_{d; 2} \cdots U^\dagger_{d; M-1} U^\dagger_{d; M}, \label{eq_009}
\ee
with $U_{c;m}=\exp(-iH_{c; m}\tau)$ and $U^\dagger_{d;m}=\exp(iH_{d; m}\tau)$.

Next we introduce an objective function,
\be
{\mathcal F}_U = \|U^\dagger_d U_c - U_{\mathrm{CZ}}\|^{2},  \label{eq_010}
\ee
where the Euclidean norm of matrix $R$ is defined as
$\|R\| = \sqrt{\mathrm{Tr}\{R^{\dag}R\}}$. The discretization of
the $Z$-control pulse determines that this function is fully
dependent on the pulse sequence, i.e.,
$\mathcal F_U \equiv \mathcal F_U(\mu_{A; 1}, \mu_{A; 2}, \cdots, \mu_{A; M})$.
For each $m$-th amplitude, the gradient of the objective function, $k_{U,m}=\partial\mathcal{F}_U/\partial\mu_{A;m}$, is given by
\be
k_{U,m} 
&=& -2 \mathrm{Re} \Tr\left\{U_{\mathrm{CZ}}U_{d}^{\dag} \frac{\partial U_{c}}{\partial\mu_{A;m}}\right\} -2 \mathrm{Re} \Tr\left\{U_{c}^{\dag} \frac{\partial U_{d}}{\partial\mu_{A;m}}U_{\mathrm{CZ}} \right\},
\label{eq_011}
\ee
where Re stands for the real part. By neglecting the commutation terms, the two partial derivatives are approximated as
$\partial U_{c}/\partial\mu_{A;m} \approx -i\tau U_{c} Q_{c; m}$ and $\partial U_{d}/\partial\mu_{A;m} \approx -i\tau U_{d} Q_{d; m}$.
Here we introduce two abbreviations, $Q_{c; m}= R^\dagger_{c; m} n_A R_{c; m}$ with $R_{c; m}=U_{c; m}U_{c; m-1}\cdots U_{c; 1}$
and $Q_{d; m}=R^\dagger_{d; m} n_A R_{d; m}$ with $R_{d; m}=U_{d; m}U_{d; m-1}\cdots U_{d; 1}$. The gradient in Eq.~(\ref{eq_011}) is simplified
to be
\be
k_{U;m}
&\approx& -2\tau \mathrm{Im}\Tr\{ U_{\mathrm{CZ}} U_{d}^{\dag}U_{c} Q_{c;m} \}
-2\tau\mathrm{Im}\Tr\{U_{\mathrm{CZ}} U^\dagger_{c}U_{d}Q_{d;m}\},
\label{eq_012}
\ee
where Im stands for the imaginary part.

The optimization of the CZ gate is given by the minimization of
the objective function, which leads to an array of $M$ equations,
\be
k_{U; m=1, 2\cdots, M}=0.
\label{eq_012a}
\ee
However, this optimization condition is nearly impossible to be solved analytically
and we apply the GRAPE method based on an iteration approach~\cite{RebingWuPRA18}.
The protocol begins with an initial guess of the pulse sequence,
${\boldsymbol\mu}^{(0)}_{A}=\{\mu^{(0)}_{A; 1}, \cdots, \mu^{(0)}_{A; M}\}$.
At each $l$-th step, we numerically calculate the gradient sequence,
${\boldsymbol k}^{(l)}=\{k^{(l)}_{1}, \cdots, k^{(l)}_{M}\}$,
and update the pulse sequence using a linear propagation,
\be
\mu^{(l+1)}_{A;m} = \mu^{(l)}_{A;m} + \alpha k^{(l)}_{m},
\label{eq_013}
\ee
where the learning rate $\alpha$ is an empirical constant.
Through a series of iteration steps,
\be
\cdots\rightarrow{\boldsymbol \mu}^{(l)}\rightarrow{\boldsymbol k}^{(l)}\rightarrow{\boldsymbol \mu}^{(l+1)}\rightarrow\cdots, \no
\ee
the gradient sequence approaches very small values (${\boldsymbol k}^{(l)}\approx 0$)
and the pulse sequence is nearly invariant
(${\boldsymbol \mu}^{(l+1)}_A\approx{\boldsymbol \mu}^{(l)}_A$).
The optimization of the $Z$-control pulse 
is thus achieved. There are a few issues to be emphasized in this protocol:
(1) Equation~(\ref{eq_013}) is a simple updating strategy and more complicated ones are allowed.
(2) The high-dimensional optimization is highly dependent on the initial guess.
(3) Although the optimized pulse sequence can be obtained through a pure numerical computation,
the experimental deviation in the pulse shape requires a data-driven approach to minimize the
influence of the residual errors in the control line~\cite{RebingWuPRA18}.
Accordingly, the gradient in Eq.~(\ref{eq_012}) is replaced by
\be
k^{(l)}_{U;m} &\approx& -2\tau \mathrm{Im}\Tr\{ U_{\mathrm{CZ}} U^{(l)}_\mathrm{exp} Q^{(l)}_{c;m} \}
 - 2\tau\mathrm{Im}\Tr\{ U_{\mathrm{CZ}} (U^{(l)}_\mathrm{exp})^\dagger Q^{(l)}_{d;m} \}.
\label{eq_014}
\ee
Here $Q^{(l)}_{c; m}$ and $Q^{(l)}_{d; m}$ are numerically
calculated using the input ${\boldsymbol \mu}^{(l)}_A$-sequence,
while $U^{(l)}_\mathrm{exp}=(U^{(l)}_d)^\dagger U^{(l)}_c$ is experimentally
estimated.

\begin{figure}[tp]
\centering
\includegraphics[width=0.5\columnwidth]{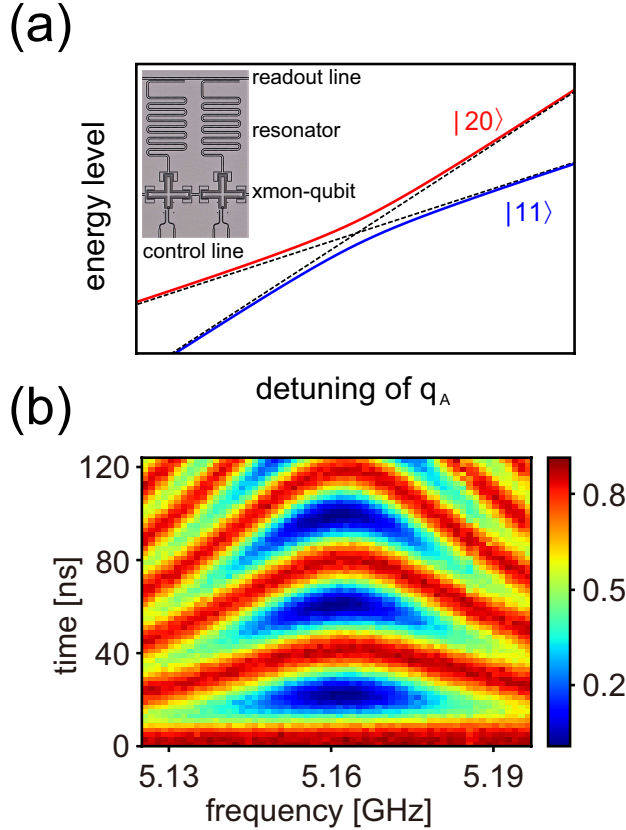}
\caption{(a) The blue and red solid lines present an avoided crossing for the states $|11\rangle$ and $|20\rangle$, 
while the two crossed dashed lines denote energy levels with a zero coupling. Inset is an optical micrograph composed of two coupled X-shaped transmon qubits.
(b) The probability  $P_{11}$ of the state $|11\rangle$ versus the target qubit frequency and the swap time.}
\label{fig_n02}
\end{figure}

\section{Experimental Implementation of Protocol I}
\label{sec3}

\subsection{Setup}
\label{sec3a}

In the inset of Fig.~\ref{fig_n02}(a), we show an image of two coupled X-shaped
transmon qubits~\cite{BarendsPRL13,wthNJP18}, in which four arms of each qubit are connected to the readout
resonator, the $XY$-control line, the $Z$-control line, and the neighboring qubit.
Each qubit is biased at an operation frequency through its $Z$-control line. In our
experiment, the two qubits are initially biased at $\omega_A/2\pi=5.458$ GHz
and $\omega_B/2\pi=4.919$ GHz, while their qubit anharmonicities
are $\Delta_A/2\pi=-242.1$ MHz and $\Delta_B/2\pi=-258.8$ MHz.
In our CZ gate operation, the fast tuning of $\omega_A$ is implemented
by an external pulse $\mu_{A}(t)$ through the $Z$-control line.
At designated operation points, the relaxation times are $T_{A; 1}=15.3~\mu$s
and $T_{B; 1}=27.9~\mu$s and the pure dephasing times are $T_{A; \phi}=13.8~\mu$s
and $T_{B; \phi}=42.7~\mu$s. Microwave drive pulses are transported
through the $XY$-control lines to control the single-qubit gate.
In the qubit state measurement, a measure pulse is transported through the
readout line, interacts with read-out resonators, and outputs a
read-out signal for the later amplification and data-collection. The frequencies
of two read-out resonators are $\omega_{A; r}/2\pi=6.462$ GHz
and $\omega_{B; r}/2\pi=6.443$ GHz. The read-out fidelities of the ground state
$|0\rangle$ and the excited state $|1\rangle$ are $F_A^{0}=97.8\%$
and $F_A^{1}=93.7\%$ for qubit $A$, and $F_B^{0}=95.2\%$
and $F_B^{1}=90.4\%$ for qubit $B$. The $Z$ line cross talk are
simultaneously calibrated and corrected, with residue coefficients below $0.2\%$.

The two-qubit CZ gate mainly depends on the $Z$-control pulse, which could be
distorted with rising or falling edges due to a filtering effect.
The $Z$ line response is calibrated and corrected, with a method similar to that in
Ref.~\cite{BarendsNat14}. The deconvolution parameters are thereafter embedded
in the underlying program to automatically correct imperfections of
the $Z$ line response. To verify this correction, we measure a swap
spectrum between the $|11\rangle$ and $|20\rangle$ states.
Figure~\ref{fig_n01}(b) shows the measured probability
$P_{11}$ as a result of the detuning time and the detuned frequency of qubit $A$.
A typical chevron pattern is observed, which confirms the reliability of the $Z$ line correction.
This chevron pattern also enables  precise extraction of two experimental parameters,
the coupling strength $g$ and the resonant frequency between $|11\rangle$ and
$|20\rangle$ states.

\subsection{Experimental results}
\label{sec3b}

In this subsection, we present our experimental result of an optimal CZ gate pulse
under the data-driven GRAPE protocol I.
The initial guess $\mu^{(0)}_A(0\le t\le T)$  is selected to follow a flattop waveform as
\be
\mu^{(0)}_A(t)=\frac{\Gamma}{2}\left[\mathrm{Erf}\left(4\sqrt{\ln{2}}(t\sigma^{-1}-1)\right)
- \mathrm{Erf}\left(4\sqrt{\ln{2}}(t\sigma^{-1}+1-T\sigma^{-1})\right)\right],
\label{eq_015}
\ee
where $\mathrm{Erf}$ denotes the error function. To demonstrate the capability of the GRAPE protocol,
we manually deviate the amplitude and the operation time away from their ideal values under a square
pulse shape. In particular, the parameters in our experiment are set as $\Gamma/2\pi= -290.6$ MHz,
$T=50$ ns, and $\sigma=4$ ns.

\begin{figure}[tp]
\centering
 \includegraphics[width=0.8\textwidth]{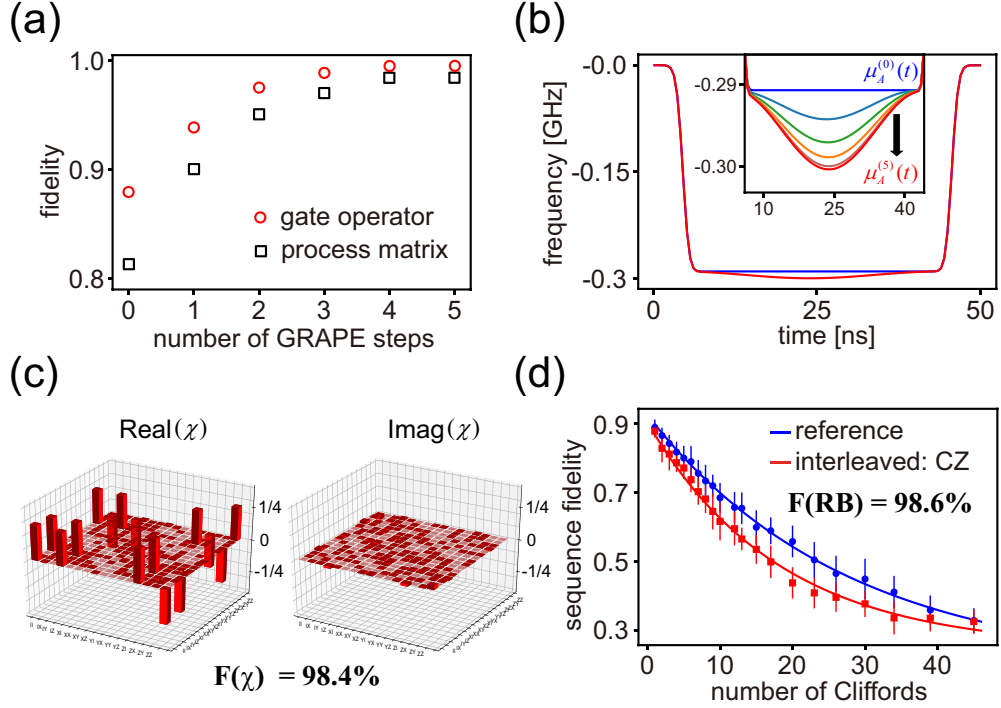}
\caption{The experimental results of the data-driven GRAPE protocol I.
(a) The fidelities of the process matrix $F(\chi^{(l)})$ (black squares) and the fitted gate operator $F(U^{(l)}_\mathrm{exp})$ (red circles) as functions of the iteration step $l$.
(b) The pulse shape modification through the iteration procedure. The initial flattop waveform (blue) and the optimal pulse $\mu^{(5)}_A(t)$ (red) are shown.
In the inset, the resonance region of six pulses ($\mu^{(0)}_A(t)\rightarrow\mu^{(5)}_A(t)$ from top to bottom) are enlarged.
(c) The QPT measurement of the $\chi$-matrix subject to the optimal CZ gate.
(d) The sequence fidelities (reference in blue and interleaved in red) versus the number of the Clifford gates, where the CZ gate is implemented by
the optimal $Z$ pulse. For each result, an error bar of the standard deviation is shown together with the average value.}
\label{fig_n03}
\end{figure}

To quantify the fidelity of the initial CZ gate, we perform a QPT
measurement. As shown in Fig.~\ref{fig_n01}(a), each qubit ($A$ or $B$)
is prepared at an initial state from the set of
$\{{|0\rangle,|1\rangle,(|0\rangle \pm |1\rangle )/\sqrt{2},(|0\rangle \pm i|1\rangle )/\sqrt{2}}\}$,
which is created by a ground state qubit subject to the pulses of $\{I, X,\pm Y/2, \mp X/2 \}$.
The total 36 initial states are inspected in a single QPT measurement.
For each initial state, its final state after the CZ gate operation
is measured by the QST and calibrated with
the read-out fidelities to eliminate the state preparation and measurement errors~\cite{wthPRAPP19}.
If the initial density matrix is $\rho(0)$,
the output counterpart $\rho^{(0)}(T)$ is in general expanded into
\be
        \rho^{(0)}(T) = \sum_{mn} \chi^{(0)}_{mn} \tilde{E}_m\rho(0) \tilde{E}_n^\dagger,
\label{eq_016}
\ee
where $\{\tilde{E}_m\}$ is a complete set of two-qubit operators.
As a full description of the gate operation, the $\chi^{(0)}$-matrix ($\chi^{(0)}=\{\chi^{(0)}_{mn}\}$)
is numerically determined using the QST data of $\rho^{(0)}(T)$
from all the 36 initial states. Then we calculate the process fidelity
using $F(\chi^{(0)}) = \mathrm{Tr}\{(\chi^{(0)})^\dagger \chi_\mathrm{ideal}\}$,
where $\chi_\mathrm{ideal}$ is the ideal matrix~\cite{wthPRAPP19}.
As shown in Fig.~\ref{fig_n03}(a), the initial flattop form $\mu^{(0)}_A(t)$ leads to the gate
fidelity at $F(\chi^{(0)}) = 81.4\%$, which suggests an improvement necessary in the $Z$-control pulse.

To fulfill the GRAPE protocol I, we need to employ the gate operator $U^{(0)}_\mathrm{exp}$. 
In our experiment, the Powell method~\cite{Powell64} is utilized to extract an estimation
of $U^{(0)}_\mathrm{exp}$ from 
the $\chi^{(0)}$-matrix. 
The fidelity of the initial gate operator,
$F(U^{(0)}_\mathrm{exp})=\mathrm{Tr}\{(U^{(0)}_\mathrm{exp})^\dagger U_\mathrm{CZ}\}/4$,
is estimated at 88.2\%. Here we must emphasize that the operator $U_\mathrm{exp}$
is less reliable than the process matrix $\chi$  due to a possible overfitting of the former in a smaller space.
However, an analytical relation between $\chi$  and $\mu_A(t)$ is extremely difficult to be extracted so that a direct
optimization of the $\chi$-matrix 
is highly inefficient. We then use the estimated
gate operator $U^{(0)}_\mathrm{exp}$ to calculate the gradient sequence ${\boldsymbol k}^{(1)}_U=\{k^{(1)}_{U; 1},   \cdots,   k^{(1)}_{U; M}\}$ and
the pulse sequence ${\boldsymbol \mu}^{(1)}_A=\{\mu^{(1)}_{A; 1}, \cdots, \mu^{(1)}_{A; M}\}$.
The above iteration procedure is repeated upto convergence.  At each $l$-th iteration step,
the time step of discretization is set to be $\tau=0.5$ ns considering the resolution limit of our arbitrary
waveform generator (AWG)~\cite{XiangPRApp20}.
The learning rate is empirically set to be $\alpha = 0.03$ GHz$^2$.
After the experimental measurement of the $\chi^{(l)}$-matrix and the numerical estimation of $U^{(l)}_\mathrm{exp}$,
the discrete pulse sequence ${\boldsymbol \mu}^{(l+1)}_A$ is calculated  by Eq.~(\ref{eq_014})
and interpolated to be a continuous function $\mu^{(l+1)}_A(t)$, which is sent to the AWG for the $(l+1)$-th gate operation.

The behavior of the iteration procedure is summarized in Figs.~\ref{fig_n03}(a)-(b).
A bump is created in the resonance region of $\mu_A(t)$ where the pulse amplitude
is enhanced to compensate an insufficient phase accumulation in
$|11\rangle$. In the first three steps 
of the iteration procedure, the gate fidelity is quickly improved from
$F(\chi^{(0)}) = 81.4\%$ to $F(\chi^{(3)}) = 97.0\%$, in parallel with 
$F(U^{(0)}_\mathrm{exp})=88.2\%\rightarrow F(U^{(3)}_\mathrm{exp})=99.3\%$.
Afterwards, the shape modification of the $Z$-control pulse slows down and
the same for the improvement of the fidelities. The pulse $\mu^{(5)}_A(t)$
after five iteration steps leads to the gate fidelity at $F(\chi^{(5)}) = 98.4\%$
and the fitted operator fidelity at $F(U^{(5)}_\mathrm{exp})=99.9\%$. Notice that $F(U^{(l)}_\mathrm{exp})$
is consistently larger than $F(\chi^{(l)})$ due to the overfitting of $U^{(l)}_\mathrm{exp}$.

For simplicity, we terminate the iteration procedure and choose $\mu^{(5)}_A(t)$ to be our optimal pulse of the CZ gate.
Figure~\ref{fig_n03}(c) presents a detailed structure
of the $\chi^{(5)}$-matrix. Through an expansion over
$\{\tilde{E}_m=\tilde{E}_A\tilde{E}_B\}$ with $\tilde{E}_A, \tilde{E}_B\!\in\!\{I, X, Y, Z\}$,
the real part of the $\chi^{(5)}$-matrix is located at the four corners
and the imaginary part of the $\chi^{(5)}$-matrix is close to zero,
in an excellent agreement with the ideal result.
Next we implement a RB measurement to quantify the gate fidelity alternatively~\cite{BarendsNat14}.
Two qubits are prepared at $|00\ra$ and driven by a sequence of
$n$ random Clifford gates, and the ground state population ($P_{00}$) is measured after a recovery gate.
For such a reference sequence, an interleaved one is formed by adding a CZ gate after each Clifford gate. 
The average populations $\bar{P}_{00}(n)$ and $\bar{P}^\pr_{00}(n)$ over 30 reference and interleaved sequences respectively
are also plotted in Fig.~\ref{fig_n03}(d).
Both populations are fitted by 
$f(n)=a p^n+b$ with $f(n)=\bar{P}_{00}(n)$ and $\bar{P}^\pr_{00}(n)$~\cite{BarendsNat14}.
The fidelity of the interleaved CZ gate is defined as $F(\mathrm{RB})\!=\!1\!-\!(3/4)(1\!-\!p_\mathrm{CZ}/p_\mathrm{ref})$, where
the fitting parameters $p_\mathrm{ref}$ and $p_\mathrm{CZ}$ refer to the reference
and interleaved sequences, respectively~\cite{BarendsNat14}. For our optimal pulse $\mu^{(5)}(t)$,
the RB fidelity is estimated at $F(\mathrm{RB}|\mu^{(5)}(t))=98.6\%$.

\section{Data-Drive GRAPE Protocol II}
\label{sec4}

In our first data-driven GRAPE protocol, the objective function is designed
to optimize the CZ gate operator $U_\mathrm{exp}$, which is
indirectly obtained by the Powell algorithm acting on the $\chi$-matrix. This
approach requires the QPT measurement over 36 initial states at each iteration
step, leading to a relatively slow optimization process. In this section, we design an
alternative protocol based on the optimization of a target density matrix,
as shown by the schematic diagram in Fig.~\ref{fig_n01}(b).

With a weak quantum dissipation, the time evolution of the density matrix
$\rho(t)$ is  described by the Lindbald master equation as~\cite{MukamelBook}
\be
\dot{\rho}(t)&=&-i[H(t),\rho(t)]\no \\
&&+\sum_{i=A, B}\sum_{j=1}^2\left(L_{i;j}\rho(t) L_{i;j}^\dag - \frac{1}{2}\{\rho(t),L_{i;j}^\dag L_{i;j}\}\right),
\label{eq_017}
\ee
where $L_{i=A,B; j=1,2}$ are the Lindblad operators.
For each qubit ($i=A, B$), the Lindblad operators,
$L_{i; 1}$ and $L_{i; 2}$, refer to the relaxation and the pure dephasing,
respectively. In the Liouville superspace~\cite{MukamelBook}, Eq.~(\ref{eq_017}) is  formally rewritten as
\be
\dot{\rho}(t)= -i\mathcal L(t)\rho(t),
\label{eq_018}
\ee
where $\mathcal L(t)$ is the Liouville superoperator including the influence
of both the system Hamiltonian and the bath-induced dissipation (see Appendix~\ref{appA}).
Since the realization of the CZ gate involves the coupled Hamiltonian
$H_c(t)=H_0(g)+H_\mathrm{ext}(t)$ and the decoupled one $H_d(t)=H_0(g=0)+H_\mathrm{ext}(t)$,
two Liouville superoperators are needed in our derivation, i.e.,
$H_c(t)\rightarrow \mathcal L_c(t)$ and $H_d(t)\rightarrow \mathcal L_d(t)$.
However, the auxiliary time evolution over $H_d(t)$ is performed by the phase
measurement so that the dissipation is ignored.

Following the approach in Sec.~\ref{sec2}, the operation time $T$ is
discretized into $M$ segments. The two Liouville superoperators become
$\mathcal L_{c}(t)\Rightarrow\{\mathcal L_{c; 1}, \mathcal L_{c; 2}, \cdots, \mathcal L_{c; M}\}$
and $\mathcal L_{d}(t)\Rightarrow\{\mathcal L_{d; 1}, \mathcal L_{d; 2}, \cdots, \mathcal L_{d; M}\}$,
where the terms at each $m$-th segment are dependent
on the external pulse $\mu_{A; m}$, i.e.,
$\mathcal L_{c; m} \equiv \mathcal L_{c; m}(\mu_{A; m})$ and
$\mathcal L_{d; m} \equiv \mathcal L_{d; m}(\mu_{A; m})$. The partial
time evolution superoperators in the Liouville space are defined as
$\mathcal U_{c; m}=\exp(-i\mathcal L_{c; m} \tau)$ and
$\mathcal U_{d; m}=\exp(-i\mathcal L_{d; m} \tau)$. For a given initial
state $\rho(0)$, the output density matrix at time $T$ is written as
%\be
$\rho (T) = \mathcal U^{-1}_d \mathcal U_c \rho (0)$
%\label{eq_019}
%\ee
with
\be
\mathcal U_c &=& \mathcal U_{c; M} \mathcal U_{c; M-1}\cdots \mathcal U_{c; 2} \mathcal U_{c; 1}, \label{eq_020}\\
\mathcal U^{-1}_d &=& \mathcal U^{-1}_{d; 1} \mathcal U^{-1}_{d; 2}\cdots \mathcal U^{-1}_{d; M-1} \mathcal U^{-1}_{d; M}.
\label{eq_021}
\ee
The reliability of the CZ gate can be described by the deviation
between the real output state $\rho(T)$ and the ideal one
$\rho_\mathrm{ideal}$, which leads to an objective function,
\be
\mathcal F_\rho &=& \|\rho(T)-\rho_\mathrm{ideal}\|^2 \no \\
&\approx& 2-2\Tr\{\rho(T)\rho_\mathrm{ideal}\}.
\label{eq_022}
\ee
The optimization criterion is given by the zero gradients,
$k_{\rho; m=1,\cdots,M}=\partial \mathcal F_{\rho}/\partial \mu_{A; m}=0$,
achieved by the GRAPE approach. With an initial guess
${\boldsymbol \mu}^{(0)}_A(t)$, we also update the external field using
a linear propagation over gradients, i.e.,
$\mu^{(l+1)}_{A; m} = \mu^{(l)}_{A; m}+\alpha k^{(l)}_{\rho; m}$.
This iteration is finished when the pulse sequence is converged
by ${\boldsymbol \mu}^{(l+1)}_A\approx {\boldsymbol \mu}^{(l)}_A$
and ${\boldsymbol k}^{(l)}_\rho\approx 0$.
In detail, each gradient is given by
\be
k^{(l)}_{\rho; m}&\approx& 2i\tau \rho^\dagger_{\mathrm{ideal}}\mathcal Q^{(l)}_{c; m} \rho^{(l)}_\mathrm{exp}(T)
-2i\tau \rho^\dagger_{\mathrm{ideal}}\mathcal Q^{(l)}_{d; m}\rho^{(l)}_\mathrm{exp}(T),
\label{eq_023}
\ee
with $\mathcal Q^{(l)}_{c; m}= (\mathcal R^{(l)}_{c; m})^{-1} \mathcal P_{A} \mathcal R^{(l)}_{c; m}$
and $\mathcal Q^{(l)}_{d; m}= (\mathcal R^{(l)}_{d; m})^{-1} \mathcal P_{A} \mathcal R^{(l)}_{d; m}$.
Here $\mathcal P_A = [n_A, \cdots]$ denotes the commutator over the number operator $n_A$, and the two partial time evolution superoperators are
$\mathcal{R}^{(l)}_{c;m} = (\mathcal{U}^{(l)}_{c;m+1})^{-1} \cdots (\mathcal{U}^{(l)}_{c;M})^{-1} \mathcal{U}^{(l)}_{d}$ and
$\mathcal{R}^{(l)}_{d;m} = \mathcal{U}^{(l)}_{d;m} \cdots \mathcal{U}^{(l)}_{d;1}$.
In practice, $\mathcal Q^{(l)}_{c; m}$ and $\mathcal Q^{(l)}_{d; m}$
are numerically calculated using the $l$-th pulse sequence ${\boldsymbol \mu}^{(l)}$,
while the output density matrix $\rho^{(l)}_\mathrm{exp}(T)$ are experimentally determined by the QST measurement.
An average over multiple output states 
can  improve the applicability of this GRAPE protocol.

\begin{figure}[tp]
\centering
 \includegraphics[width=0.8\textwidth]{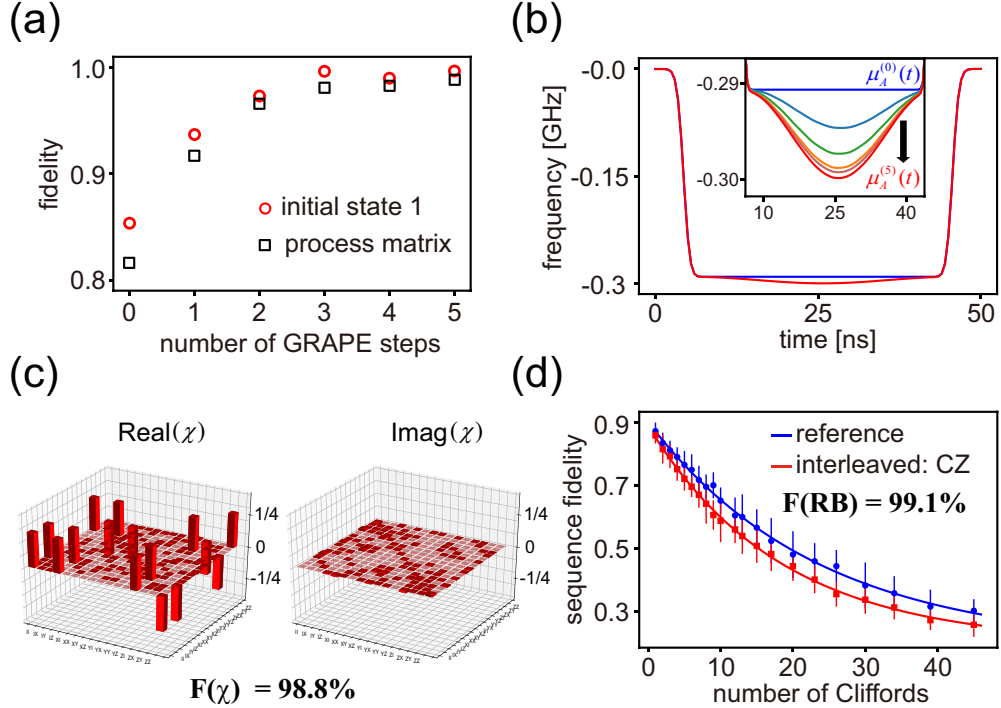}
\caption{The experimental results of the data-driven GRAPE protocol II.
(a) The  fidelities of an output density matrix $F(\rho^{(l)}_{\mathrm{exp}}(T)|\varphi_1(0))$ (red circles) and the process matrix $F(\chi^{(l)})$ (black squares) as functions of the iteration step $l$.
(b) The pulse shape modification through the iteration procedure.
(c) The QPT measurement of the $\chi$-matrix subject to the optimal CZ gate.
(d) The sequence fidelities under the optimal $Z$ pulse in the RB measurement. The legends in (b)-(d) are the same as those in Figs.~\ref{fig_n03}(b)-(d).}
\label{fig_n04}
\end{figure}

\section{Experimental Implementation of Protocol II}
\label{sec5}

In this section, we present the experimental result of an optimal CZ gate under the data-driven GRAPE protocol II.
The same flattop waveform with the same parameters as in Sec.~\ref{sec3b} is selected for the initial guess $\mu^{(0)}_A(t)$.
Four specific initial states, $\varphi_{1,2}(0)=(|0\rangle\pm|1\rangle)\otimes(|0\rangle\pm i|1\rangle)/2$
and $\varphi_{3,4}(0)=(|0\rangle\pm i|1\rangle)\otimes(|0\rangle\pm |1\rangle)/2$,
are considered in the optimization procedure. These initial states are created by the pulses of $\{\pm Y/2, \mp X/2\}$  and $\{\mp X/2, \pm Y/2\}$
applying to the two ground-state qubits. For the initial $Z$-control pulse, the fidelities of the four output density
matrices, $F(\rho^{(0)}_\mathrm{exp}(T)) = \mathrm{Tr}\{\rho^{(0)}_\mathrm{exp}(T)\rho_\mathrm{ideal}\}$, are in the range of $81.5\%\sim85.3\%$,
consistent with the gate fidelity $F(\chi^{(0)})=81.4\%$. For each output density matrix, we calculate the gradient sequence
${\boldsymbol k}^{(1)}_\rho$ using Eq.~(\ref{eq_023}) and obtain an updated pulse sequence ${\boldsymbol \mu}^{(1)}_A$.
The time step is $\tau=0.5$ ns and the learning rate is $\alpha = 0.1$ GHz$^2$. The average of four pulse sequences
are interpolated to generate a continuous form $\mu^{(1)}_A(t)$ for the subsequent gate operation. This procedure is
then repeated until being terminated at $\mu^{(5)}_A(t)$.

The evolution of the iteration process is shown in Figs.~\ref{fig_n04}(a)-(b).
The CZ gate is quickly improved in the first three steps
and then gradually approaches an optimal result. For example, the fidelity of the output density matrix
evolved from $\varphi_{1}(0)$ becomes $F(\rho^{(3)}_\mathrm{exp}(T))= 99.6\%$ after three iteration steps
and is stabilized above 99\% thereafter (see Fig.~\ref{fig_n04}(a)). For clarity, we perform the QPT measurement
for each $l$-th $Z$-control pulse $\mu^{(l)}_A(t)$ despite the fact the $\chi$-matrix is unnecessary in the protocol II.
As shown in Fig.~\ref{fig_n04}(b), the gate fidelity is improved as
$F(\chi^{(0)})=81.4\%\rightarrow F(\chi^{(3)})=98.0\%\rightarrow F(\chi^{(5)})=98.8\%$.
Similar to the behavior in the protocol I, the shape modification mainly occurs in the resonance region of the $Z$-control
pulse, in which an additional bump is created for the sufficient phase accumulation.

The pulse $\mu^{(5)}_A(t)$ obtained after five iteration steps is treated as the second optimal $Z$-control pulse
of the CZ gate. In Fig.~\ref{fig_n04}(c), we present the corresponding structure of the $\chi^{(5)}$-matrix,
which agrees excellently with an ideal one. In Fig.~\ref{fig_n04}(d), we present the result of the RB measurement
using $\mu^{(5)}_A(t)$ as the interleaved CZ gate. Following the data analysis method in Sec.~\ref{sec3},
we obtain the RB fidelity of the second optimal CZ gate at $F(\mathrm{RB}|\mu^{(5)}_A(t))=99.1\%$,
in comparable to the result from the GRAPE protocol I.

\section{Summary}
\label{sec6}

In this paper, we experimentally implement the optimization of the two-qubit
CZ gate based on two different data-driven GRAPE protocols.
These two protocols are designed to
minimize two different objective
functions based on the fitted gate operator and a target output density matrix, respectively.
Following a feedback-control mechanism, the key step in each protocol utilizes
mixed information of the input $Z$-control pulse and the experimental measurements (the QPT and the QST)
to numerically calculate a gradient sequence, which leads to the subsequent $Z$-control pulse.
A well fabricated quantum device of superconducting X-shaped
transmon qubits is used for the realization of these two GRAPE protocols.
For both protocols, we quickly obtain the optimal $Z$-control pulses around 5 iteration steps.
The resulted two CZ gates are confirmed to yield high fidelities in the QPT measurement ($98.4\%$ and $98.8\%$)
and the RB measurement ($98.6\%$ and $99.1\%$).

The main advantage of the GRAPE algorithm is its efficiency
in the convergence speed, especially by optimizing the density matrices in the second protocol.
In the previous RB-based Nelder-Mead algorithm, the pulse sequence with different number of Clifford gates
should be explored and each case requires a large number of random sequences, despite that only
the ground state population is be measured. Nevertheless,
the search speed of the Nedler-Mead algorithm is intrinsically slower than that
of the GRAPE algorithm. As a result, the Nedler-Mead is more suitable for the
parameter optimization under a fixed waveform while the GRAPE for a pulse sequence
over a fixed operation time. In general, there always exist many, sometime a huge number of,
possibilities in the problem of high-dimensional optimization.
Our experimental results show that the gate fidelities from two different data-driven GRAPE
protocols are close to each other and comparable with those from previous RB-based Nelder-Mead
experiments. Overall, various algorithms
compose a comprehensive strategy for the optimization of the CZ gate.

\section*{Acknowledgements}
The work reported here was supported by
the National Key Research and Development Program of China (Grant No. 2019YFA0308602, No. 2016YFA0301700),
the National Natural Science Foundation of China (Grants No. 12074336, No. 11934010, No. 11775129),
the Fundamental Research Funds for the Central Universities in China, and the Anhui
Initiative in Quantum Information Technologies (Grant No. AHY080000). Y.Y. acknowledge
the funding support from Tencent Corporation. This work was partially conducted at the
University of Science and Technology of the China Center
for Micro- and Nanoscale Research and Fabrication.

\appendix
\section{Liouville Superoperators}
\label{appA}

In this Appendix, we summarize the Liouville superoperators in the Lindblad equation.
The superoperator for the commutator of the system Hamiltonian $H(t)$ is
\be
[\mathcal L_{\mathrm{sys}}(t)]_{k_1l_1,k_2l_2} = [H(t)]_{k_1, k_2} \delta_{l_1, l_2}- [H(t)]_{l_2, l_1} \delta_{k_1, k_2},
\label{eq_A01}
\ee
For each qubit, the superoperator for the population relaxation part is
\be
[\mathcal{L}_{\mathrm{relax}}]_{k_1l_1,k_2l_2} = \frac{i}{T_1}\left[\sqrt{k_2l_2} \delta_{k_1+1,k_2}\delta_{l_1+1,l_2}-\frac{1}{2}(k_1+l_1)\delta_{k_1,k_2}\delta_{l_1,l_2}\right],
\label{eq_A02}
\ee
and the superoperator for the pure dephasing part is
\be
[\mathcal{L}_{\mathrm{deph}}]_{k_1l_1,k_2l_2} = -\frac{i}{T_{\phi}}(k_1-l_1)^2\delta_{k_1,k_2}\delta_{l_1,l_2}.
\label{eq_A03}
\ee

\end{document}

\title{Supplementary Material for ``Optimization of Controlled-$Z$ Gate with Data-driven Gradient Ascent Pulse Engineering in a Superconducting Qubit System"}
\author{Zhiwen Zong}
   \thanks{These authors have contributed equally to this work.}
 \affiliation{Zhejiang Province Key Laboratory of Quantum Technology and Device, Department of Physics, Zhejiang University, Hangzhou, 310027, China}
 \author{Zhenhai Sun}
    \thanks{These authors have contributed equally to this work.}
 \affiliation{Zhejiang Province Key Laboratory of Quantum Technology and Device, Department of Physics, Zhejiang University, Hangzhou, 310027, China}
   \author{Zhangjingzi Dong}
 \affiliation{Zhejiang Province Key Laboratory of Quantum Technology and Device, Department of Physics, Zhejiang University, Hangzhou, 310027, China}
   \author{Chongxin Run}
 \affiliation{Zhejiang Province Key Laboratory of Quantum Technology and Device, Department of Physics, Zhejiang University, Hangzhou, 310027, China}
\author{Liang Xiang}
 \affiliation{Zhejiang Province Key Laboratory of Quantum Technology and Device, Department of Physics, Zhejiang University, Hangzhou, 310027, China}
  \author{Ze Zhan}
 \affiliation{Zhejiang Province Key Laboratory of Quantum Technology and Device, Department of Physics, Zhejiang University, Hangzhou, 310027, China}
  \author{Qianlong Wang}
 \affiliation{Zhejiang Province Key Laboratory of Quantum Technology and Device, Department of Physics, Zhejiang University, Hangzhou, 310027, China}
   \author{Ying Fei}
 \affiliation{Zhejiang Province Key Laboratory of Quantum Technology and Device, Department of Physics, Zhejiang University, Hangzhou, 310027, China}
   \author{Yaozu Wu}
 \affiliation{Zhejiang Province Key Laboratory of Quantum Technology and Device, Department of Physics, Zhejiang University, Hangzhou, 310027, China}
   \author{Wenyan Jin}
 \affiliation{Zhejiang Province Key Laboratory of Quantum Technology and Device, Department of Physics, Zhejiang University, Hangzhou, 310027, China}
    \author{Cong Xiao}
 \affiliation{Zhejiang Province Key Laboratory of Quantum Technology and Device, Department of Physics, Zhejiang University, Hangzhou, 310027, China}
 \author{Zhilong Jia}
 \affiliation{Key Laboratory of Quantum Information, University of Science and Technology of China, Hefei, 230026, China}
 \author{Peng Duan}
 \affiliation{Key Laboratory of Quantum Information, University of Science and Technology of China, Hefei, 230026, China}
 \author{Jianlan Wu }
 \email{jianlanwu@zju.edu.cn}
 \affiliation{Zhejiang Province Key Laboratory of Quantum Technology and Device, Department of Physics, Zhejiang University, Hangzhou, 310027, China}
 \author{Yi Yin}
 \email{yiyin@zju.edu.cn}
 \affiliation{Zhejiang Province Key Laboratory of Quantum Technology and Device, Department of Physics, Zhejiang University, Hangzhou, 310027, China}
 %\affiliation{Collaborative Innovation Center of Advanced Microstructures, Nanjing, 210093, China}
 \author{Guoping Guo}
 \email{gpguo@ustc.edu.cn}
 \affiliation{Key Laboratory of Quantum Information, University of Science and Technology of China, Hefei, 230026, China}
 \affiliation{Origin Quantum Computing, Hefei, 230026, China}

\maketitle

\section{non-adiabatic CZ gate}
\label{sec1}

The Hamiltonian of our two-qubit system %under an external field
can be efficiently defined in the Hilbert space of
$\{|00\rangle, |10\rangle, |01\rangle, |11\rangle, |20\rangle\}$~\cite{wthPRAPP19}. In detail, the Hamiltonian
is written as
\be
H_c(t) &=& 0 |00\rangle\langle 00|+[\omega_A+\mu_A(t)]|10\rangle\langle 10|+\omega_B |01\rangle\langle 01| \no \\
&& + [\omega_A+\mu_A(t)+\omega_B]|11\rangle\langle 11| + [2(\omega_A+\mu_A(t))+\Delta_A]|20\rangle\langle 20| \no \\
&& + \sqrt{2}g \left[|11\rangle\langle20|+|20\rangle\langle 11|\right],
\ee
where the coupling $g$ between $|10\rangle$ and $|01\rangle$ is ignored.
Under the resonance condition of $\mu_A(t)=\mu_{A;r}=\omega_B-\omega_A-\Delta_A$, the above Hamiltonian is simplified to be
\be
H_c &=& 0 |00\rangle\langle 00|+[\omega_A+\mu_{A;r}]|10\rangle\langle 10|+\omega_B |01\rangle\langle 01| \no \\
&& + [\omega_A+\mu_{A;r}+\omega_B]\left[|11\rangle\langle 11| + |20\rangle\langle 20|\right] \no \\
&& + \sqrt{2}g \left[|11\rangle\langle20|+|20\rangle\langle 11|\right].
\label{eq_S002}
\ee
On the other hand, the auxiliary Hamiltonian without the inter-qubit coupling ($g=0$) is explicitly given by
\be
H_d &=&0 |00\rangle\langle 00|+[\omega_A+\mu_{A;r}]|10\rangle\langle 10|+\omega_B |01\rangle\langle 01| \no \\
&& + [\omega_A+\mu_{A;r}+\omega_B]\left[|11\rangle\langle 11| + |20\rangle\langle 20|\right].
\label{eq_S003}
\ee

Next we construct the time evolution operators, $U_c=\exp(-iH_c T)$ and $U_d =\exp(-iH_d T)$. For the  Hamiltonian $H_c$
defined in Eq.~(\ref{eq_S002}), the coupling only exists between $|11\rangle$ and $20\rangle$.
In this two-dimensional (2D) subspace, we introduce two operators, $I_2=|11\rangle\langle11|+|20\rangle\langle20|$
and $X_2=|11\rangle\langle20|+|20\rangle\langle11|$ so that the $2\times2$  Hamiltonian is rewritten as
\be
H_{\mathrm{sub}} = [\omega_A+\mu_{A;r}+\omega_B]I_2+ \sqrt{2}g X_2.
\ee
The corresponding time evolution operator over the time lap $T$ is given by
\be
U_{\mathrm{sub}} = e^{-i(\phi_A+\phi_B)} \exp(-i\sqrt{2}gT X_2),
\label{eq_S005}
\ee
with $\phi_A=(\omega_A+\mu_{A;r})T$ and $\phi_B=\omega_B T$. After a straightforward derivation, the second term on the right hand side
of Eq.~(\ref{eq_S005}) is simplified to be
\be
\exp(-i\sqrt{2}gT X_2) = \cos(\sqrt{2}gT) I_2-i \sin(\sqrt{2}gT) X_2.
\label{eq_S006}
\ee
Under the condition of a half period, $T=\pi/\sqrt{2}g$, this $2\times2$ time evolution operator is
\be
U_{\mathrm{sub}} = -e^{-i(\phi_A+\phi_B)}I_2 =-e^{-i(\phi_A+\phi_B)} [|11\rangle\langle 11|+|20\rangle\langle 20|],
\ee
and the total time evolution operator is
\be
U_c %&=& \exp(i H_c T) \no \\
&=& |00\rangle\langle 00| + e^{-i\phi_A}|10\rangle \langle 10| + e^{-i \phi_B} |01\rangle\langle 01|-e^{-i(\phi_A+\phi_B)}[|11\rangle\langle 11|+|20\rangle\langle 20|].
\ee
On the other hand, the time evolution operator without the inter-qubit coupling is
\be
U_d %&=& \exp(i H_d T) \no \\
&=& |00\rangle\langle 00| + e^{-i\phi_A}|10\rangle \langle 10| + e^{-i \phi_B} |01\rangle\langle 01|+e^{-i(\phi_A+\phi_B)}[|11\rangle\langle 11|+|20\rangle\langle 20|].
\ee
The combination of these two operators leads to
\be
U_\mathrm{CZ} &=&  U^\dagger_d U_c \no \\
&=& |00\rangle\langle 00| + |10\rangle \langle 10| + |01\rangle\langle 01|-|11\rangle\langle 11|-|20\rangle\langle 20|.
\label{eq_S010}
\ee
If the initial state is in the subspace of $\{|00\rangle, |10\rangle, |01\rangle, |11\rangle\}$, the last term on the right hand side of Eq.~(\ref{eq_S010})
can be ignored and this time evolution operator forms an ideal CZ gate.

\section{Derivation of the Gradient in Protocol I}
\label{sec2}

Due to the residual errors in the control line during the time evolution, the simple square-shaped pulse ($\mu_A(0<t<T)=\mu_{A; r}$)
cannot produce a high-fidelity CZ gate. Therefore, we need to modify the shape of $\mu_A(t)$. The two Hamiltonians,
$H_c(t)$ and $H_d(t)$, are time-variant and the corresponding time evolution operators are changed to
$U_c(T)=T_+ \exp[-i \int_0^T H_c(t) dt]$ and $U_d(T)=T_+ \exp[-i \int_0^T H_d(t) dt]$, where $T_+$ stands for the forward
time ordering operator. For convenience, we divide the time lap $T$ into $M$ segments ($\tau=T/M$) and the external field $\mu_A(t)$
is discretized into
\be
\mu_A(t) \Rightarrow \{\mu_{A; 1}=\mu_A(t=0), \mu_{A; 2}=\mu_A(\tau), \cdots, \mu_{A; M}=\mu_A((M-1)\tau) \}.
\ee
The two Hamiltonians become
\be
H_{c; m=1, 2, \cdots, M}&=& 0 |00\rangle\langle 00|+[\omega_A+\mu_{A;m}]|10\rangle\langle 10|+\omega_B |01\rangle\langle 01| \no \\
&& + [\omega_A+\mu_{A;m}+\omega_B]|11\rangle\langle 11| + [2(\omega_A+\mu_{A;m})+\Delta_A]|20\rangle\langle 20| \no \\
&& + \sqrt{2}g \left[|11\rangle\langle20|+|20\rangle\langle 11|\right],
\ee
and
\be
H_{d; m=1, 2, \cdots, M}&=& 0 |00\rangle\langle 00|+[\omega_A+\mu_{A;m}]|10\rangle\langle 10|+\omega_B |01\rangle\langle 01| \no \\
&& + [\omega_A+\mu_{A;m}+\omega_B]|11\rangle\langle 11| + [2(\omega_A+\mu_{A;m})+\Delta_A]|20\rangle\langle 20|.
\ee
The two time evolution operators become
\be
U_c &=& U_{c; M}U_{c; M-1}\cdots U_{c;2}U_{c;1}, \no \\
U_d &=& U_{d; M}U_{d; M-1}\cdots U_{d;2}U_{d;1},
\ee
with $U_{c; m}=\exp(-i H_{c; m}\tau)$ and $U_{d; m}=\exp(-i H_{d; m} \tau)$.
The combined gate operator, $U=U^\dagger_d U_c$, however deviates from the ideal CZ gate $U_\mathrm{CZ}$.

In our first GRAPE procedure, the objective function is defined as~\cite{RebingWuPRA18}
\be
\mathcal{F}_U &=& \|U_{d}^{\dag}U_{c} - U_{\mathrm{CZ}}\|^{2} \no \\
&=& 2-\mathrm{Tr}\{ U_{\mathrm{CZ}}U_{d}^{\dag}U_{c}\}-\mathrm{Tr}\{U_{c}^{\dag}U_{d}U_{\mathrm{CZ}}\}.
\ee
Due to the discretization of the external field $\mu_A(t)$, $\mathcal{F}_U$ is a function of all the pulse
amplitudes, i.e., $\mathcal{F}_U\equiv \mathcal{F}_U(\mu_{A;1}, \mu_{A; 2}, \cdots, \mu_{A; M})$. For each $m$th
amplitude $\mu_{A; m}$, the gradient of the objective function is
\be
k_{U,m} %&=&\frac{\partial\mathcal{F}_U}{\partial\mu_{A;m}} \no \\
&=& -2 \mathrm{Re} \Tr\left\{U_{\mathrm{CZ}}U_{d}^{\dag} \frac{\partial U_{c}}{\partial\mu_{A;m}}\right\} -2 \mathrm{Re} \Tr\left\{U_{c}^{\dag} \frac{\partial U_{d}}{\partial\mu_{A;m}}U_{\mathrm{CZ}} \right\}.
\label{eq_S016}
\ee
The optimization condition is thus given by $\partial \mathcal{F}_U/\partial \mu_{A; m}=0$ for $m=1, 2, \cdots, M$.
To fulfill this condition, we further expand the gradient in Eq.~(\ref{eq_S016}).
The partial derivative of $U_{c}$ is given by
\be
\frac{\partial U_{c}}{\partial\mu_{A;m}}
&=& U_{c; M}...U_{c; m+1} \frac{\partial U_{c;m}}{\partial\mu_{A;m}} U_{c; m-1}...U_{c; 1}
\label{eq_S017}
\ee
To calculate $\partial U_{c; m}/\partial \mu_{A; m}$, we need to expand the time evolution operator $U_{c; m}$ into
\be
U_{c; m}= I-i\tau H_{c; m} -\tau^2 H^2_{c; m}/2 -\cdots,
\ee
and apply the result
\be
\frac{\partial H^n_{c; m}}{\partial \mu_{A; m}} = \frac{\partial H_{c; m}}{\partial \mu_{A; m}}H^{n-1}_{c; m}+H_{c; m}\frac{\partial H_{c; m}}{\partial \mu_{A; m}}H^{n-2}_{c; m}+\cdots.
\ee
In general, $\partial H_{c; m}/\partial \mu_{A; m}$ and $H_{c; m}$ do not commute with each other. To perform
a practical computation, we however take an acceptable approximation,
\be
\frac{ \partial H^n_{c; m}}{\partial \mu_{A; m}} \approx  n\frac{\partial H_{c; m}}{\partial \mu_{A; m}}H^{n-1}_{c; m}.
\ee
As a result, Eq.~(\ref{eq_S017}) is simplified to be
\be
\frac{\partial U_{c}}{\partial\mu_{A;m}}
&\approx& -i\tau U_{c; M}...U_{c; m+1} \frac{\partial H_{c; m}}{\partial \mu_{A; m}} U_{c; m} U_{c; m-1}...U_{c; 1}
\label{eq_S021}
\ee
With the introduction of $\partial H_{c; m}/\partial \mu_{A; m}=n_A$ and $R_{c; m}=U_{c; m}U_{c; m-1}\cdots U_{c; 1}$,
Eq.~(\ref{eq_S021}) is organized into
\be
\frac{\partial U_{c}}{\partial\mu_{A;m}}
&\approx& -i\tau U_c R^+_{c; m} n_A R_{c; m} = -i\tau U_c Q_{c; m},
\label{eq_S022}
\ee
with $Q_{c; m}= R^\dagger_{c; m}n_A R_{c; m}$. The same derivation is applied to the partial derivative of $U_d$, giving
\be
\frac{\partial U_{d}}{\partial\mu_{A;m}}
&\approx& -i\tau U_c Q_{d; m},
\label{eq_S023}
\ee
with $Q_{d; m}=R^\dagger_{d; m}n_A R_{d; m}$ and $R_{d; m}=U_{d; m}U_{d; m-1}\cdots U_{d; 1}$. By substituting Eqs.~(\ref{eq_S022})
and (\ref{eq_S023}) into Eq.~(\ref{eq_S016}), we obtain
\be
k_{U;m}
&\approx& -i\tau \Tr\left\{Q_{c;m}U^\dagger_{c} U_{d} U_{\mathrm{CZ}}-U_{\mathrm{CZ}}U_{d}^{\dag}U_cQ_{c;m}
+ U_{\mathrm{CZ}} Q_{d;m}U^\dagger_d U_{c}- U_{c}^{\dag} U_d Q_{d; m}U_{\mathrm{CZ}}  \right\}\no \\
&=& -2\tau \mathrm{Im}\Tr\{ U_{\mathrm{CZ}} U_{d}^{\dag}U_{c} Q_{c;m} \}
-2\tau\mathrm{Im}\Tr\{U_{\mathrm{CZ}} U^\dagger_{c}U_{d}Q_{d;m}\},
\label{eq_S024}
\ee
where Im stands for the imaginary part.
In experiment, the practical gate $U$ can be estimated using the Powell algorithm over the QPT data so that
Eq.~(\ref{eq_S024}) is finally rewritten as
\be
k_{U;m} \approx -2\tau \mathrm{Im}\Tr\{ U_{\mathrm{CZ}} U^{(l)}_\mathrm{exp} Q^{(l)}_{c;m} \}
 - 2\tau\mathrm{Im}\Tr\{ U_{\mathrm{CZ}} (U^{(l)}_\mathrm{exp})^\dagger Q^{(l)}_{d;m} \}.
\label{eq_S025}
\ee

\section{Estimation of the Gate Operator with the Powell Method}

In experiment, the behavior of a gate is fully described by the process matrix $\chi$, which can be obtained through the measurement of quantum processing tomography (QPT).
Due to quantum dissipation and other sources of errors, the experimental determination $\chi_{\mathrm{exp}}$ always deviates from an ideal description so that
it is nearly impossible to calculate the gate operator $U$  exactly. Instead, we can apply a fitting approach to extract the best estimation $U_{\mathrm{exp}}$,
which is used as experimental information in our data-driven GRAPE protocol I.

In practice, we choose the Powell algorithm for a relatively fast estimation of $U_{\mathrm{exp}}$~\cite{Powell64}. For a given gate operator, a $N$-dimensional parameter space $\vec{X}$
is defined according to all the $N$ independent matrix elements, i.e., $\vec{X}\equiv\vec{X}(\{U_{ij}\})$. The objective function is the square of a distance,
$\mathcal F=\|\chi(U)-\chi_{\mathrm{exp}}\|^{2} =\|\chi(X)-\chi_{\mathrm{exp}}\|^{2}$, between the extracted process matrix and the experimental measurement.
The Powell method starts with an initial guess of $\vec{X}^{(0)}_0$. At each $l$-th iteration step, there exist
the total $N$ pre-determined directions, $\{\vec{h}^{(l)}_1, \cdots, \vec{h}^{(l)}_N\}$, in the parameter space. At the follow-up $k(\le N)$-th sub-step,
we begin with the position $\vec{X}^{(l)}_{k-1}$ and search an optimal position $\vec{X}^{(l)}_k$ for the one-dimensional minimization
of $\mathcal F$ along the $\vec{h}^{(l)}_k$-direction. After the total $N$ sub-steps, we obtain a new direction,
$\vec{h}^{(l)}_{N+1}=\vec{X}^{(l)}_{N+1}-\vec{X}^{(l)}_0$, which may be used to replace one direction (e.g., $\vec{h}^{(l)}_1$)
in $\{\vec{h}^{(l)}_k\}$. This updated set is used as the direction set $\{\vec{h}^{(l+1)}_k\}$ for the next $(l+1)$-th iteration step
with the starting position at $\vec{X}^{(l+1)}_0=\vec{X}^{(l)}_{N+1}$. The whole iteration process is terminated when the objective function reaches its smallest value.

\section{Lindblad Master Equation and Its Representation in the Liouville Space}
\label{sec3}

The second GRAPE protocol in our paper is based on the optimization of a target density matrix. With the consideration
of a weak bath-induced dissipation, the time evolution of the density matrix is governed by the Lindblad master equation~\cite{MukamelBook},
\be
\dot{\rho}(t) = -i[H(t),\rho(t)]+\sum_s \left(L_s \rho(t) L_s^\dagger-\frac{1}{2} \{\rho(t), L^\dag_s L_s\}\right)
\label{eq_S026}
\ee
where $H(t)$ is the total Hamiltonian and $\{L_s\}$ is the set of Lindbad operators for relaxation and decoherence.
For two arbitrary operators $A$ and $B$ in the Hilbert space, $[A,B] = AB-BA$ and $\{A,B\}=AB+BA$ define their
commutator and anti-commutator, respectively.

To facilitate the derivation and the numerical calculation, we introduce the concepts of the Liouville space and
the Liouville superoperators. In the Hilbert space defined by the basis set of $\{|k_1\rangle\}$, the density matrix is expanded as
\be
\rho = \sum_{k_1,l_1}\rho_{k_1,l_1}|k_1\rangle\langle l_1|.
\label{eq_S027}
\ee
In the Liouville space, this matrix is converted into a vector form,
\be
\rho = \sum_{k_1,l_1}\rho_{k_1,l_1} |k_1,l_1\rangle\rangle,
\label{eq_S028}
\ee
which is built through a new basis set of $\{|k_1,l_1\rangle\rangle\}$ under the one-to-one mapping, $|k_1,l_1\rangle\rangle\Leftrightarrow |k_1\rangle\langle l_1|$.
We introduce the complex conjugate term,
$|k_1,l_1\rangle\rangle^\dagger=\langle\langle k_1,l_1| \Leftrightarrow |l_1\rangle\langle k_1|$, which gives rise to the inner product
\be
\langle\langle k_1,l_1| k_1^\pr, l_1^\pr\rangle\rangle \equiv \Tr\{|l_1\rangle\langle k_1| k_1^\pr\rangle\langle l_1^\pr| \} = \delta_{k_1, k_1^\pr} \delta_{l_1, l_1^\pr},
\label{eq_S029}
\ee
and the outer product $|k_1,l_1\rangle\rangle\langle\langle k_2,l_2|$. The identity superoperator (matrix) $\mathcal I$ in Liouville space
is expanded as
\be
\mathcal I = \sum_{k_1,l_1} |k_1,l_1\rangle\rangle \langle\langle k_1,l_1|.
\label{eq_S030}
\ee
Then, we construct an arbitrary Liouville superoperator $\mathcal L$, which follows a matrix form as
\be
\mathcal L = \sum_{k_1,l_1,k_2,l_2} \mathcal L_{k_1,l_1;k_2,l_2} |k_1,l_1\rangle\rangle \langle\langle k_2,l_2|,
\ee
with $\mathcal L_{k_1,l_1;k_2,l_2}=\langle\langle k_1,l_1|\mathcal L|k_2,l_2\rangle\rangle$.
To illustrate the application of the Liouville superoperators, we take the two examples. \\
\noindent {\bf (I) The commutator
of the system Hamiltonian.} For a general $N$-dimensional Hilbert space, the system Hamiltonian is
expanded as $H(t)=\sum_{k_1,l_1}[H(t)]_{k_1,l_1}|k_1\rangle\langle l_1|$. In the Hilbert space, the commutator is written as
\be
H(t)\rho -\rho H(t) &=& \sum_{k_1,l_1,k_2} \left([H(t)]_{k_1, k_2} \rho_{k_2, l_1}-\rho_{k_1, k_2} [H(t)]_{k_2, l_1}\right) |k_1\rangle\langle l_1| \no \\
&=& \sum_{k_1,l_1,k_2,l_2} \left([H(t)]_{k_1, k_2} \delta_{l_1, l_2}- [H(t)]_{l_2, l_1} \delta_{k_1, k_2}\right)\rho_{k_2, l_2} |k_1\rangle\langle l_1|.
\ee
Following Eqs.~(\ref{eq_S028}) and (\ref{eq_S029}), the above equation is rewritten in the Liouville space as
\be
\mathcal L_\mathrm{sys} \rho &=& H(t)\rho -\rho H(t) \no \\
&=&\left[\sum_{k_1,l_1,k_2,l_2}   \left([H(t)]_{k_1, k_2} \delta_{l_1, l_2}- [H(t)]_{l_2, l_1} \delta_{k_1, k_2}\right)|k_1,l_1\rangle\rangle\langle\langle k_2,l_2|\right]
\left[\sum_{k^\pr, l^\pr} \rho_{k^\pr, l^\pr} |k^\pr, l^\pr\rangle\rangle\right].
\ee
Thus, the commutator of the system Hamiltonian follows a matrix form as
\be
\mathcal L_\mathrm{sys} =\sum_{k_1,l_1,k_2,l_2}   \left([H(t)]_{k_1, k_2} \delta_{l_1, l_2}- [H(t)]_{l_2, l_1} \delta_{k_1, k_2}\right)|k_1,l_1\rangle\rangle\langle\langle k_2,l_2|
\label{eq_S034}
\ee
{\bf (II) The dissipation due to the Lindblad operator $L_s$.} The dissipation part of the Lindbald equation in Eq.~(\ref{eq_S026}) is expanded as
\be
\mathcal L_s \rho &=& L_s \rho L^\dagger_s-\frac{1}{2} \{\rho, L^\dagger_s L_s\} \no \\
&=& \sum_{k_1,l_1,k_2,l_2} \left[L_{s; k_1, k_2}\rho_{k_2,l_2} L^\dagger_{s; l_2, l_1}-\frac{1}{2}\rho_{k_1, k_2}(L^\dagger_{s}L_{s})_{k_2, l_1}-\frac{1}{2}(L^\dagger_s L_s)_{k_1, k_2}\rho_{k_2, l_1}\right]|k_1\rangle\langle l_1| \no\\
&\Rightarrow& \sum_{k_1,l_1,k_2,l_2} \left[L_{s; k_1, k_2} L^\dagger_{s; l_2, l_1}-\frac{1}{2}(L^\dagger_{s}L_{s})_{l_2, l_1}\delta_{k_1, k_2}-\frac{1}{2}(L^\dagger_s L_s)_{k_1, k_2}\delta_{l_2, l_1}\right]\rho_{k_2,l_2} |k_1,l_1\rangle\rangle.
\label{eq_S035}
\ee
The corresponding Liouville superoperator is organized into
\be
\mathcal{L}_{s}=\sum_{k_1,l_1;k_2,l_2} \left[L_{s; k_1,k_2} L^\dagger_{s; l_2, l_1}-\frac{1}{2}(L^\dagger_{s}L_{s})_{l_2, l_1}\delta_{k_1,k_2}-\frac{1}{2}(L^\dagger_s L_s)_{k_1,k_2}\delta_{l_2, l_1}\right]|k_1,l_1\rangle\rangle\langle\langle k_2,l_2|.
\ee
For the relaxation part of a single qubit, the Lindblad operator is given by
\be
L_{\mathrm{relax}}= \sum_{j} \sqrt{\frac{j}{T_1}}|j-1\rangle\langle j|
\ee
which leads to the Liouville superoperator,
\be
\mathcal{L}_\mathrm{relax} = \sum_{k_1,l_1;k_2,l_2} \frac{i}{T_1}\left[\sqrt{k_2l_2} \delta_{k_1+1,k_2}\delta_{l_1+1,l_2}-\frac{1}{2}(k_1+l_1)\delta_{k_1,k_2}\delta_{l_1,l_2}\right] |k_1,l_1\rangle\rangle\langle\langle k_2,l_2|
\label{eq_S038}
\ee
For the pure dephasing part of a single qubit, the Lindblad operator is given by
\be
L_\mathrm{deph} = \sum_j \sqrt{\frac{2}{T_\phi}}j|j\rangle\langle j|
\ee
which leads to the Liouville superoperator,
\be
\mathcal{L}_\mathrm{deph} = \sum_{k_1,l_1;k_2,l_2} -\frac{i}{T_{\phi}}(k_1-l_1)^2\delta_{k_1,k_2}\delta_{l_1,l_2} |k_1,l_1\rangle\rangle\langle\langle k_2,l_2|.
\label{eq_S040}
\ee
Overall, the Liouvile superoperator governing the dissipation of a single qubit is written as
\be
\mathcal{L}_\mathrm{diss} & = \sum_{k_1,l_1;k_2,l_2} i \Big[\frac{1}{T_1}\sqrt{k_2l_2}\delta_{k_1+1,k_2}\delta_{l_1+1,l_2}-\no\\
&\Big(\frac{(k_1-l_1)^2}{T_{\phi}}+\frac{(k_1+l_1)}{2T_1}\Big)\delta_{k_1,k_2}\delta_{l_1,l_2}\Big]|k_1,l_1\rangle\rangle\langle\langle k_2,l_2|
\label{eq_S041}
\ee
By substituting the detailed forms of the Liouville superoperators into the Lindblad equation, we obtain a matrix equation form for the time evolution of the density matrix, given by
\be
\dot{\rho}(t) = -i \mathcal L(t)\rho(t),
\label{eq_S042}
\ee
where the total Liouville superoperator $\mathcal L(t)$ is constructed from the system part $\mathcal L_\mathrm{sys}(t)$ and the dissipation part $\mathcal L_\mathrm{diss}$,
i.e., $\mathcal L(t) = \mathcal L_\mathrm{sys}(t)+\mathcal L_\mathrm{diss}$. Equation~(\ref{eq_S042}) is formally solved as
\be
\rho(T) = \mathcal U(T) \rho(0)~~~\mathrm{with}~~~\mathcal U(T) = T_+ \exp\left[-i \int_0^T \mathcal L(t) dt\right] .
\label{eq_S043}
\ee
Here $\mathcal U(T)$ is a superoperator governing the time evolution of the density matrix in the Liouville space.

\section{Derivation of the Gradient in Protocol II}
\label{sec5}

Equations~(\ref{eq_S042}) and (\ref{eq_S043}) serve as the starting point of our second GRAPE protocol. For the two-qubit system in our experiment, we can
either consider a 5-dimensional
Hilbert space of $\{|00\rangle, |10\rangle, |01\rangle, |11\rangle, |20\rangle\}$ or a more comprehensive 9-dimensional space of $\{|i_A(=0, 1, 2) i_B(=0, 1, 2)\rangle\}$.
Similar to the treatment in Protocol I, we consider two system Hamiltonians, $H_c(t)=H_0(g)+H_\mathrm{ext}(\mu_A(t))$ and $H_d(t)=H_0(g=0)+H_\mathrm{ext}(\mu_A(t))$.
The two Liouville superoperators $\mathcal L_c(t)$ and $\mathcal L_d(t)$ are constructed accordingly. However, the auxiliary operation over $H_d(t)$ is obtained effectively
by measuring the dynamic phases $\phi_A(T)$ and $\phi_B(T)$. As a result, $\mathcal L_c(t)$ includes the dissipation part while $\mathcal L_d(t)$ does not.
Furthermore, the discretization of the external field $\mu_A(t)\Rightarrow\{\mu_{A; 1}, \mu_{A; 2}, \cdots, \mu_{A; M}\}$ leads to two sets of Liouville superoperators,
\be
\mathcal L_c(t) &\Rightarrow& \{\mathcal L_{c; 1}, \mathcal L_{c; 2}, \cdots, \mathcal L_{c; M}\},  \\
\mathcal L_d(t) &\Rightarrow& \{\mathcal L_{d; 1}, \mathcal L_{d; 2}, \cdots, \mathcal L_{d; M}\}.
\ee
The two time evolution superoperators are given by
\be
\mathcal{U}_c &= \mathcal{U}_{c;M}\mathcal{U}_{c;M-1} \cdots \mathcal{U}_{c;2}\mathcal{U}_{c;1} \label{eq_S046} \\
\mathcal{U}_d &= \mathcal{U}_{d;M}\mathcal{U}_{d;M-1} \cdots \mathcal{U}_{d;2}\mathcal{U}_{d;1}
\label{eq_S047}
\ee
with
\be
\mathcal{U}_{c;m} &= \exp[-i\mathcal{L}_{c;m}\tau], \label{eq_S048} \\
\mathcal{U}_{d;m} &= \exp[-i\mathcal{L}_{d;m}\tau] \label{eq_S049}.
\ee
and $\tau=T/M$.

In the second GRAPE protocol, we may choose a specific initial state, e.g., $\varphi(0)=(|0\rangle + |1\rangle)\otimes(|0\rangle + i|1\rangle)/2$
and $\rho(0)=\varphi(0)\varphi^\dagger(0)$, and check whether the final state $\rho(T)=\mathcal U^{-1}_d \mathcal U_c \rho(0)$ agrees with the ideal result
$\rho_\mathrm{ideal}=U_\mathrm{CZ}\rho(0)U^\dagger_\mathrm{UZ}$. %In our real experimental, we select two initial states to reduced errors.
In the Hilbert space, we introduce the objective function
\be
\mathcal{F}_\rho &=& \|\rho_{\mathrm{f}} - \rho_{\mathrm{ideal}}\|^{2} \no \\
&=& \Tr\{[\rho(T)-\rho_\mathrm{ideal}]^+[\rho(T)-\rho_\mathrm{ideal}]\} \no \\
&\approx& 2-2\Tr\{\rho_\mathrm{ideal}\rho(T)\},
\label{eq_S050}
\ee
The derivation in Eq.~(\ref{eq_S050}) uses the conditions, $\rho^\dagger(T)=\rho(T)$, $\rho^\dagger_\mathrm{ideal}=\rho_\mathrm{ideal}$,
and $\Tr\{\rho^2_\mathrm{ideal}\}=1$. The real final state is in general a mixed state but the weak dissipation in our experiment allows a good
approximation, $\Tr\{\rho^2(T)\}\approx 1$. Equation~(\ref{eq_S050}) is next reformed into an inner product in the Liouville space as
\be
\mathcal F_\rho &\approx& 2-2\rho^\dagger_\mathrm{ideal}\rho(T) \no \\
&=& 2-2\rho^\dagger_\mathrm{ideal}\mathcal U^{-1}_d \mathcal U_c \rho(0).
\label{eq_0S51}
\ee
Similarly, the minimization of $\mathcal F_\rho$ is controlled by the conditions, $\{\partial \mathcal F_\rho/\partial \mu_{A; m=1, 2, \cdots, M}=0\}$.
In regard to the $m$-th pulse amplitude, the corresponding gradient is given by
\be
k_{\rho; m}&=&\frac{\partial {\mathcal F_\rho}}{\partial \mu_{A;m}} \no \\
&=&-2\rho^\dagger_{\mathrm{ideal}}\left[\frac{\partial\mathcal{U}_{d}^{-1}}{\partial\mu_{A;m}}\mathcal{U}_{c}+\mathcal{U}_{d}^{-1}\frac{\partial\mathcal{U}_{c}}{\partial\mu_{A;m}}\right]\rho(0).
\label{eq_S052}
\ee
Following the similar approximations,
\be
\partial \mathcal U_{c; m}/\partial \mu_{A; m} &\approx& -i\tau(\partial \mathcal L_{c; m}/\partial \mu_{A; m})\mathcal U_{c; m} \\
\partial \mathcal U^{-1}_{d; m}/\partial \mu_{A; m} &\approx& i\tau\mathcal U^{-1}_{d; m}(\partial \mathcal L_{d; m}/\partial \mu_{A; m}),
\ee
and the condition, $\partial \mathcal L_{c; m}/\partial \mu_{A; m}=\partial \mathcal L_{d; m}/\partial \mu_{A; m}=\mathcal P_A$ with $\mathcal P_A =[n_A, \cdots]$,
Eq.~(\ref{eq_S052}) is simplified to be
\be
k_{\rho; m}&\approx&2i\tau \left[\rho^\dagger_{\mathrm{ideal}}\mathcal Q_{c; m} \mathcal U^{-1}_d \mathcal U_c \rho(0)-\rho^\dagger_{\mathrm{ideal}}\mathcal Q_{d; m}\mathcal U^{-1}_d \mathcal U_c \rho(0)\right]
\label{eq_S055}
\ee
with
\be
\mathcal Q_{c; m} &=& \mathcal R^{-1}_{c; m} \mathcal P_{A} \mathcal R_{c; m},  \\
\mathcal Q_{d; m} &=& \mathcal R^{-1}_{d; m} \mathcal P_{A} \mathcal R_{d; m},
\ee
and
\be
\mathcal{R}_{c;m} &=& \mathcal{U}_{c;m+1}^{-1} \cdots \mathcal{U}_{c;M}^{-1} \mathcal{U}_{d}, \\
\mathcal{R}_{d;m} &=& \mathcal{U}_{d;m} \cdots \mathcal{U}_{d;1}.
\ee
Furthermore, if the final density matrix $\rho(T)=\mathcal U^{-1}_d \mathcal U_c \rho(0)$ is measured experimentally, the gradient in Eq.~(\ref{eq_S055}) becomes
\be
k_{\rho; m}&\approx&2i\tau \left[\rho^\dagger_{\mathrm{ideal}}\mathcal Q_{c; m} \rho_\mathrm{exp}(T)-\rho^\dagger_{\mathrm{ideal}}\mathcal Q_{d; m}\rho_\mathrm{exp}(T)\right],
\label{eq_S060}
\ee
where $\rho_\mathrm{exp}(T)$ denotes the experimental result. Notice that both terms, $\rho^\dagger_{\mathrm{ideal}}\mathcal Q_{d; m}\rho_\mathrm{exp}(T)$
and $\rho^\dagger_{\mathrm{ideal}}\mathcal Q_{c; m} \rho_\mathrm{exp}(T)$, are pure imaginary so that the gradient $k_{\rho; m}$ is guaranteed to be real.